\documentclass[amssymb,aps,prli,preprint]{revtex4}
\usepackage{latexsym,amsmath,graphics,graphicx,exscale,colordvi,color}
\begin{document}

\title{Magnetic adatom induced skyrmion-like spin texture in surface electron waves}
\author{Samir Lounis}\email{s.lounis@fz-juelich.de}
\author{Andreas Bringer}
\author{Stefan Bl\"ugel}
\affiliation{Peter Gr\"unberg Institut \& Institute for Advanced Simulation, 
Forschungszentrum J\"ulich \& JARA, D-52425 J\"ulich,
Germany}
\begin{abstract}
 When 
a foreign atom is placed on a surface of a metal, the surrounding 
sea of electrons responds screening the additional 
charge leading to oscillations or ripples.  On surfaces, those electrons are  
sometimes confined to two-dimensional surface states, whose spin-degeneracy is
lifted due to the Rashba effect arising from the spin-orbit  interaction of electrons and the inversion asymmetric environment. It is believed that at least for a single adatom scanning tunneling microscopy measurements are 
insensitive to the Rashba splitting {\it i.e.}\ no signatures in the charge oscillations will be observed. 
Resting on scattering theory, we demonstrate that, if magnetic, one single adatom is enough to visualize the presence of the Rashba effect in terms of  an induced spin-magnetization of the 
surrounding electrons exhibiting a twisted spin texture described as superposition of two skyrmionic waves
of opposite chirality. 
\end{abstract}
\maketitle
%\date{\today}

%\pacs{}

%\begin{multicols}{2}
%narrowtext

\begin{figure}%[ht!]
\begin{center}
\includegraphics*[angle=0,width=1.\linewidth]{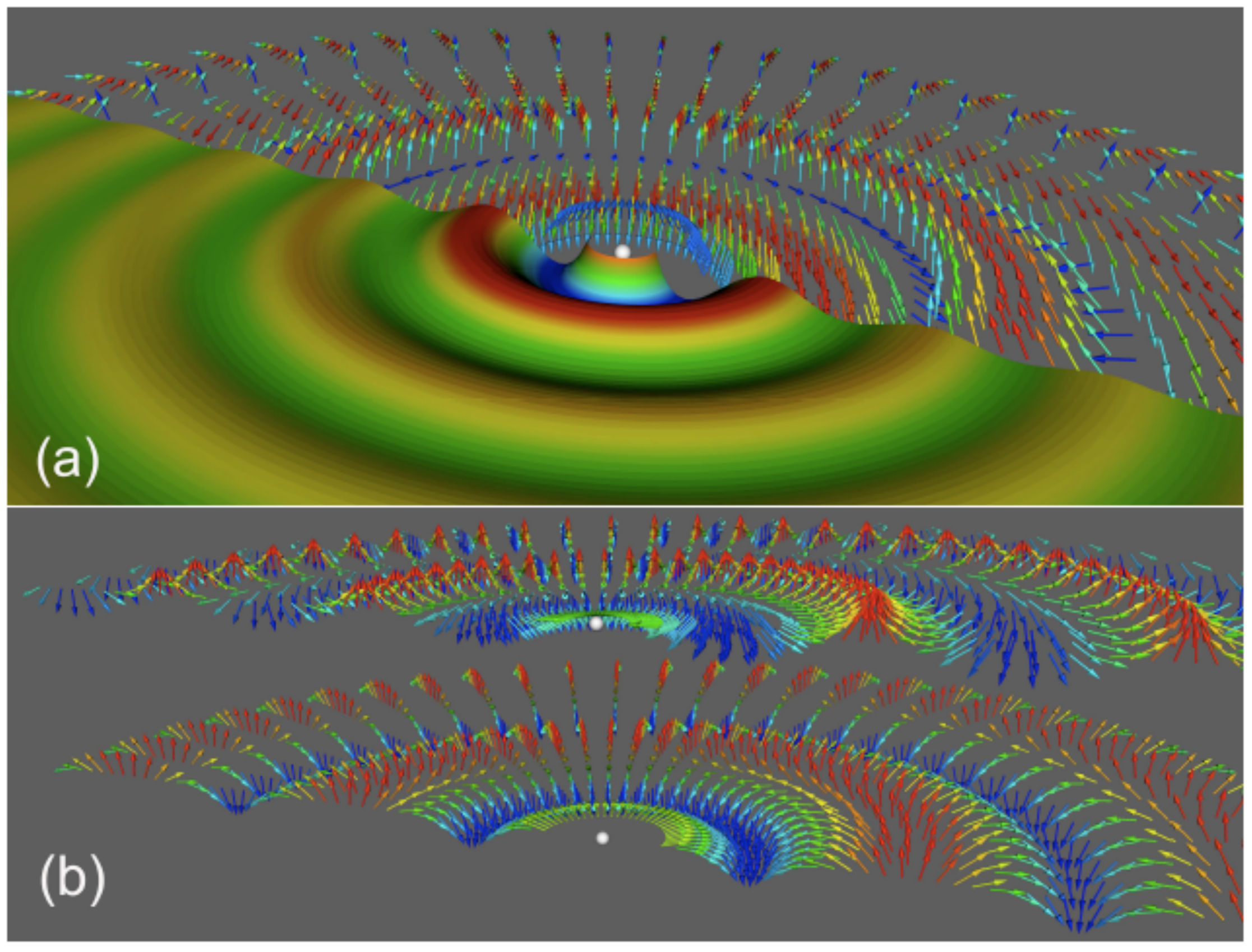}
\end{center}
\caption{Skyrmionic-like spin-texture at the Fermi energy of Au(111) surface. 
(a) Spatial visualization of the induced local density of states (LDOS)  and 
the local magnetization direction  of Au surface electrons
after scattering with an Fe adatom.  The spin texture found in (a) 
can be decomposed into a linear combination of two smoothly rotating 
skyrmionic magnetic waves with opposite vector chirality shown in (b). 
The wavelengths of LDOS oscillation, the left and right skyrmionic waves are  $\sim 18.7$~\AA, $\sim 17.3$~\AA\ and $20.3$~\AA, respectively.}
\label{skyrmion}
\end{figure}

 The lack of 
spatial inversion symmetry is the triggering  ingredient for a wide range of new phenomena that are accessible 
with state of the art experimental 
techniques~\cite{roessler,roch,heinze,bode,koroteev,ast2}. 
Angle-resolved photoemission spectroscopy (ARPES) was the first tool used by LaShell 
and coworkers~\cite{lashell} to discover a small energy  splitting in the $sp$ surface state band of the  
Au(111) surface. This splitting has been interpreted by 
the same authors as a realization of the interaction between the spin and orbital angular momentum, which promoted the
 renaissance of the 
Rashba physics. The surge of interest 
in similar effects involving the spin-orbit interaction in a structural asymmetric environment is the incitement of many additional sophisticated measurements and theoretical simulations on relevant surfaces with~\cite{hasan,balatsky,franz} and without~\cite{reinert,nicolay,henk,koroteev,ast,pascual} topologically protected surface 
states.
 Since charge oscillations
(Friedel oscillations) induced by scattering of
the surface-state electrons at a single adatom are blind
with respect to such spin-orbit effects [16], 
an alternative [17] has been proposed on the basis of a  multiple scattering study consisting on probing with the scanning tunneling microscope (STM) 
charge oscillations confined
within a corral of adatoms. 
Theoretical calculations~\cite{ngo} outline  the possibility of very complex magnetic structures. We pursue a different 
route and investigate the possibility of 
grasping information on the spin-orbit interaction at surfaces exploiting the break of  time inversion symmetry introduced 
by a magnetic adatom. As shown and discussed later, we 
discover an intriguing spin-texture in the spin-polarized electron 
gas surrounding the magnetic adatom 
(Fig.~\ref{skyrmion}). This new magnetic behavior can be very similar to the topological twists, called skyrmions~\cite{skyrme,roessler,roch,heinze,belavin}. Parts of our predictions have recently been verified in the interferences 
produced by the scattering at a MnPc molecule of the complex surface states of Bi(110) surface~\cite{pascual}. 

\begin{figure}%[ht!]
\begin{center}
\includegraphics*[width=.5\linewidth]{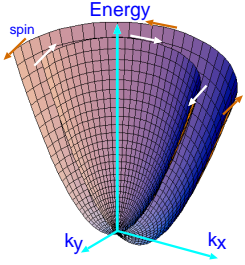}
\end{center}
\caption{Energy dispersion of spin-split surface-state electrons. When plotted with respect to the two components of the 
two-dimensional wave vector $\vec{k}$, the energy dispersion are cone-shaped.  Due to spin-orbit interaction, two spin-split parabolas are obtained and are centered around $\pm k_{so}$ (see text). 
 Arrows indicate the 
vector fields of the spin-quantization axes (or the patterns of the spin) at 
the constant energy contour. For every energy, two opposite spins have different wave vectors leading to two concentric circle with opposite spins. The effective B-field 
felt by the electrons is perpendicular to the propagation direction defined by $\vec{k}$.}
\label{host_Au}
\end{figure}

Our investigation is based on a Rashba Hamiltonian~\cite{rashba} that 
describes a two-dimensional gas of free electrons confined in the ($xy$) 
plane of a metallic surface: 
\begin{equation}
\hat{H} =\frac{\hat{p}_x^2 + \hat{p}_y^2}{2m^*} - \frac{\alpha_R}{\hbar}
(\sigma_x\hat{p}_y -\sigma_y\hat{p}_x)
\label{rashba_h}
\end{equation}
considered with respect to a zero-energy reference defined by the bottom of the dispersion curve in absence of the spin-orbit coupling. $m^*$ is the 
effective mass of the electron and $\alpha_R$ is the effective Rashba parameter, describing the strength of the effect, whose value is determined in principle by the 
atomic spin-orbit strength as well as by the degree of asymmetry of 
the wavefunction imposed by the presence of 
the surface~\cite{petersen}. Here, however, $\alpha_R$ is chosen to model the experimental dispersion relation of 
the suface state. Quite generally the gradient of the potential at the surface acts as an  electric field $\vec{E}$ normal to the surface
in the lab frame of the sample. Electrons propagating with momentum 
$\vec{k}$ across the surface experience this field in their local frame of reference 
as an effective magnetic field, $\vec{B}=\frac{\hbar}{m^*c}\vec{k}\times\vec{E}$, 
 which is the origin of the functional form for the Rashba Hamiltonian and defines a spin-quantization axis $\hat{n}(\vec{k})$ to be located in the surface
 plane normal to the wavevector $\vec{k}=(k_x,\,k_y)=k(\cos\phi,\,\sin\phi)\perp\hat{n}(\vec{k})$, and $c$ is the speed of light. 
The eigenvectors $\vert\psi_{1(2)}\rangle$ of the Hamiltonian (\ref{rashba_h}), 
 associated with the wavevectors $\vec{k}_1$ and $\vec{k}_2$, are spin-up and spin-down states,
respectively, with respect to $\hat{n}$, but are also a coherent superposition of
spin-up and -down states,
$\vert\!\psi_{1(2)}(\vec{k})\rangle=\frac{e^{ik_{1(2)}}}{\sqrt{2}}(\vert
 \uparrow\rangle -(+) i e^{i\phi}\vert\downarrow\rangle)$,
expressed by the spin functions $\vert \uparrow\rangle$ and 
$\vert \downarrow\rangle$, when 
measured with respect to the surface normal ($z$-direction). The 
eigenvalue spectrum $E_{1(2)}(\vec{k})=\frac{\hbar^2}{2m^*}(k_{1(2)}^2-k_{\mathrm{so}}^2)$
is a two spin-split cone-shaped parabolic energy dispersion curve by which 
the origins of the parabola $E_{1(2)}$ are shifted with 
respect to $k=0$ by $k_{\mathrm{so}}=\pm m^*\alpha_{R}/\hbar^2$, {\em i.e.\  } $k_{1(2)}=k+(-)k_{\mathrm{so}}$  (see Fig.~\ref{host_Au}). 
When $k_2$ changes sign from positive to 
negative, the two branches of the dispersion curves cross. 
Unless stated otherwise, we shall consider in the upcoming text 
only positive $k_2$.

The scattering of the surface-state Rashba electrons on a magnetic adatom 
deposited on a substrate is investigated using scattering theory,
which has been successfully applied in the description of 
electron scattering at adatoms~\cite{heller,eigler,manoharan}. 
This theory involves the calculation of 
Green functions, which allow an elegant treatment of the electronic properties
including the embedded adatom  by solving the Dyson equation, 
$G= G_0 +  G_0tG_0$, where all quantities are site, spin and energy dependent 
matrices. We note that the electronic properties of the adatom are inscribed 
into the scattering matrix $t$, the 
amplitude of the electron-wave scattering at the adatom.
$G_0$  is the Green function of the pure two-dimensional electron gas
corresponding to the Hamiltonian (\ref{rashba_h}) that is  
constructed using the eigenvectors $|\psi_{1(2)}\rangle$:
\begin{equation}
G_0 =
\begin{bmatrix}
% G_\mathrm{D} & \exp({-i\phi}) G_\mathrm{ND}\\
% -\exp({i\phi}) G_\mathrm{ND} & G_\mathrm{D}
 G_\mathrm{D} & e^{-i\phi} G_\mathrm{ND}\\
 -e^{i\phi} G_\mathrm{ND} & G_\mathrm{D}
\end{bmatrix} .
\end{equation}
In case the Rashba effect vanishes, the off-diagonal part of the 
Green function, $G_\mathrm{ND}$, is zero and the diagonal part, $G_\mathrm{D}$,
reverts to the Green function of the free electron surface states.
 From $G$,
quantities related to those measured by STM, such as  
the local density of states (LDOS) ~\cite{tersoff} 
$n(\vec{r};E)= -1/\pi\, \mathrm{Im Tr}_{\sigma} G(\vec{r},\vec{r}; E)$ or measured by
the spin-polarized STM (SP-STM) as the local magnetization density of states (LMDOS)~\cite{wortmann} 
$\vec{M}(\vec{r};E)=-1/\pi\,\mathrm{Im Tr}_{\sigma} \vec{\sigma}G(\vec{r},\vec{r}; E)$ 
can be calculated, where $\vec{\sigma}$ is the vector of Pauli matrices and 
Tr$_{\sigma}$ is a trace in spin space. It is convenient to express all 
quantities in cylindrical coordinates as it turns out that they depend
only on the radial distance $R$ measured from the position of the adatom.  The LMDOS is expressed by two components,  
$\vec{M}(R)=(M_z,M_r)$, the  radial 
component, $M_r$, and the one normal to the surface, $M_z$. The azimuthal
component, $M_\phi$, vanishes. 

The obtained results are general, but to strengthen our point we consider 
as an application the system of an Fe adatom on the Au(111) surface, since 
the system has been proven experimentally accessible by low-temperature 
STM and for Au the presence of a large Rashba effect has been shown experimentally~\cite{lashell}.
On such a surface, Jamneala and 
coworkers~\cite{jamneala} found that contrary to Ti, Co, and Ni, no Kondo peak 
was observed for V, Cr, Mn, and Fe adatoms meaning, that for the latter elements
indeed a local magnetic moment exists and it 
is not screened by the conduction electrons~\cite{kondo}. 
We treat Fe as an adatom with a  magnetic moment pointing along the $z$-direction perpendicular to the surface since 
our first-principles calculations~\cite{kkr} (see supplementary material) predict this to be 
the easy axis with a magnetic anisotropy energy of about $ 12$~meV.  

Hence, such a case imposes a scattering matrix that is diagonal in spin-space,
\begin{eqnarray}
t &=&\begin{bmatrix}
t_{\uparrow\uparrow}& 0 \\
0& t_{\downarrow\downarrow}
\end{bmatrix},
\end{eqnarray}
which can be related to the phase-shift 
$\delta(E)$ experienced by the incoming electron waves scattering at 
the adatom  ($t = \hbar^2/m^*[\exp(2i\delta(E))-1]$). 
 Note that a generalization to an arbitrary rotation angle of the magnetic moment will be 
straightforward using standard unitary transformations. 
For Fe, as  for Au, all majority-spin states are fully occupied 
and the scattering of majority electrons in the vicinity of the Fermi energy ($E_{\mathrm{F}}$)  is practically
zero, i.e.\ the phase-shift vanishes ($t_{\uparrow\uparrow}=0$). 
The minority-spin LDOS shows, on the contrary to Au, high values around the Fermi energy and therefore a large scattering. Thus, we 
assume a phase-shift of $\pi/2$ ($t_{\downarrow\downarrow} = -2\hbar^2/m^*$).

The parameters describing the Au(111) surface state are identical to 
those chosen by Walls and Heller~\cite{walls}: $m^* = 0.26\, m_e$,
$E_F = 0.41$~eV and $\alpha_{R} = 0.4$~eV\AA\ that correspond to a     
Fermi wavelength $\lambda_{\mathrm{F}}=\frac{2\pi}{k} \sim  37.4$~\AA\ and 
a spin rotation length $\lambda_{\mathrm{so}}=\frac{\pi}{k_{\mathrm{so}}}$ of 230.5~\AA.

After solving the Dyson equation, the induced circular LDOS oscillations,
$\Delta n(R;E)$, emanating from the adatom located at the center ($R=0$) is given by
$\Delta n=-1/\pi\, {\mathrm{Im}} [(G_\mathrm{D}G_\mathrm{D}+G_\mathrm{ND}G_\mathrm{ND})(t_{\uparrow\uparrow}+t_{\downarrow\downarrow})]$ and is
plotted in Fig.~\ref{skyrmion}(a) for all states at $E=E_{\mathrm{F}}$. 
At large distances $R$,  
the circular induced standing wave undulations, $\Delta n(R;E)$, can be expressed as $\sim  \frac{m^*}{\pi^2\hbar^2k^2R}\sqrt{k_1k_2}\cos(2kR)$. When the spin-orbit coupling is negligible, {\em i.e.}\  $k_1 \sim k_2$, we recover the conventional form of the adatom induced energy dependent  charge density oscillations $\sim \frac{\cos(2kR)}{2kR}$~\cite{eigler,heller}. 
 Regardless of the spin-orbit interaction, there is {\em only one} wave vector,  
$2k= k_1+k_2$,  that describes the oscillations,  with the corresponding wavelength at $E_F$ given by $\lambda_{F}/2 \sim 18.7$~\AA\ . Consistent to the work of  Petersen and Hedeg{\aa}rd~\cite{petersen} and 
confirmed by Walls and Heller~\cite{walls}, no signature or information 
on the Rashba effect or spin-orbit interaction of the surface atoms can be 
extracted from the LDOS. 

This picture changes fundamentally when we look at the LMDOS. Due to the presence
of a magnetic adatom ($t_{\uparrow\uparrow}\ne t_{\downarrow\downarrow}$),
the magnetization density perpendicular to the 
surface {\em and} in the surface plane  are non-zero. We find concentric
rings $M(R)$ of equal size magnetization density surrounding the Fe adatom,
with magnetization densities emanated at the center of the Fe atom wrangling 
in the $(M_z(R),\, M_r(R))$ plane (Fig.~\ref{skyrmion}(a)). 
Contrary to the LDOS, the magnetization density is non-trivially modulated by 
the Rashba effect. Analysing the asymptotic behavior of 
$M_z(R) \sim\frac{m^*}{2\pi^2 \hbar^2 k^2 R}(k_1 \cos(2k_1 R) + k_2 \cos(2k_2 R))$
as calculated from 
$M_z=-1/\pi\, {\mathrm{Im}} [(G_\mathrm{D}G_\mathrm{D}-G_\mathrm{ND}G_\mathrm{ND})(t_{\uparrow\uparrow}-t_{\downarrow\downarrow})]$)
as well as of
$M_r(R) \sim \frac{m^*}{2\pi^2\hbar^2k^2R}(k_1\sin(2k_1R)-k_2\sin(2k_2R))$
as calculated from 
$M_{r}(R) = - 2/\pi\, {\mathrm{Im}} [G_\mathrm{D}(t_{\uparrow\uparrow}-t_{\downarrow\downarrow})G_\mathrm{ND}]$,
 one finds that both wave vectors $k_1$ and $k_2$ enter nontrivially.
This observation implies the possibility of using 
a SP-STM with an out-of-plane magnetized tip to probe the $z$-projected 
 Rashba induced interferences. Any  magnetic signal detected by a 
SP-STM with an in-plane magnetized tip would 
be a clear fingerprint of the Rashba effect. If there is no Rashba effect,
i.e.\ $G_\mathrm{ND}=0$, the quantization axis is only determined by 
the adatom and the scattering electrons are not a coherent superposition 
of spin-up and down-electrons that that led to a finite in-plane spin-component.

\begin{figure}%[ht!]
\begin{center}
\includegraphics*[width=1.\linewidth]{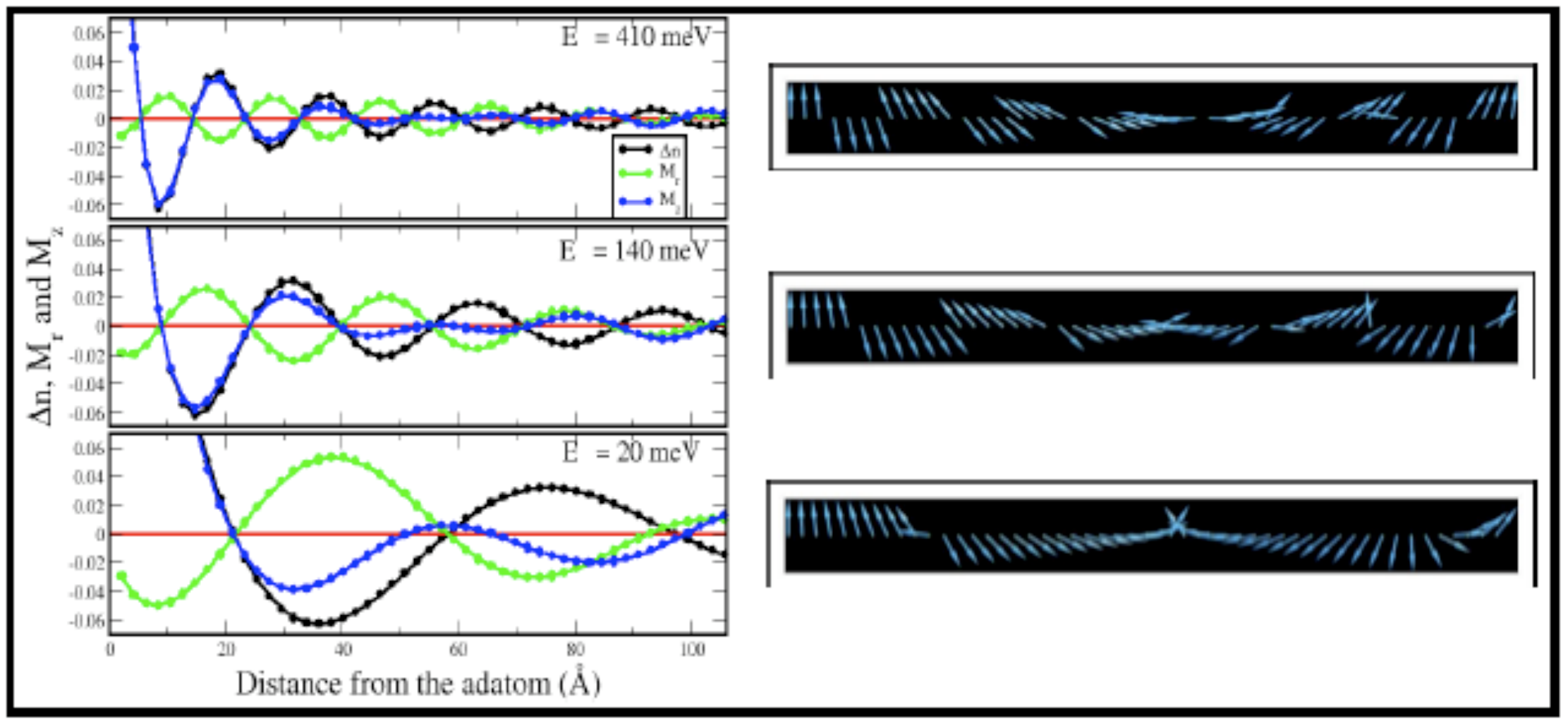}
\end{center}
\caption{ Comparison of the induced LDOS and LMDOS components. 
Due to the Rashba effect the Fe magnetic adatom induces non-trivial spin interferences: on  
the left panels are shown the radial dependence of the LDOS and LMDOS at different energies while on the right panels   
the corresponding magnetization unit vectors are depicted.}
\label{dos_yaxis}
\end{figure}

Such spin-textures are a reminder of skyrmionic magnetic 
configurations but are completely different. Indeed, with 
skyrmions, the magnetic moments can rotate smoothly from one direction at the center of 
the structure to the opposite direction. Here, however, the magnetization ripples experience 
phase-switch as well as beating as sketched  in the top-right of 
Fig.~\ref{dos_yaxis}. In order to simplify our analysis, the induced LDOS and the $z$- and radial components of the magnetization passing by the adatom position are additionally illustrated in the 
top-left of the same figure. Surprisingly, the amplitude of the $M_r$ and $M_z$ oscillations 
are of the same size with a 
phase-shift between them that depends on the strength of the Rashba coupling term. 
This can be better understood when looking at the 
asymptotic behavior. For instance, $M_z$ can be 
rewritten in terms of $k_{\mathrm{so}}$ and $k$ as $\sim \frac{m^*}{\pi^2 \hbar^2 
k^2 R} (k\cos(2kR)\cos(2k_{\mathrm{so}}R) - k_{so} \sin(2kR)\sin(2k_{\mathrm{so}}R))$. 
 At the Fermi 
energy of gold, $k_{\mathrm{so}}$ is expected to be very 
small compared to $k$. At distances much smaller than the spin-rotation length but long enough that the 
previous asymptotic behavior holds, $M_z$ simplifies more to 
$\frac{m^*}{\pi^2 \hbar^2kR}\cos(2kR)$ and oscillates similarly to the LDOS with or without spin-orbit interaction. This 
explains the wavelengths of the induced initial ripples being, as expected,  
 twice the Fermi wavevector.  
In fact one has to go far beyond the vicinity of the adatom, 
at distances of the order of the spin-rotation length, 
to see an effect achieved by the intriguing 
phase switch observed around 60~\AA.
Indeed, the new spin-rotation length corresponding to a 
full rotation from $0^o$ to $180^o$ occurs 
around $115$~\AA, which means that the 
change of sign occurs at half this value. 
 Another substrate with a stronger Rashba coupling parameter 
would impose a more important phase-shift at smaller distances.
  $M_{r}$, which is proportional to 
$-\frac{m^*}{\pi^2 \hbar^2 k^2R}(k\cos(2kR)\sin(2k_{\mathrm{so}}R)+k_{\mathrm{so}}\sin{2kR}\cos(2k_{\mathrm{so}}R))$ can 
be further simplified -- for rather short distances -- to $-\frac{2m^*k_{\mathrm{so}}}{\pi^2\hbar^2k}\cos(2kR)$, similar to the 
cosine behavior that characterizes $M_z$. 

Although the spin-texture looks very complex at first sight it can be 
understood as a linear combination of two skyrmionic waves, 
$\vec{M}=\vec{M}_{k_1}+\vec{M}_{k_2}$, with 
$\vec{M}_{k_1}(R)\propto k_1(\cos(2k_1R),-\sin(2k_1R))$ and 
$\vec{M}_{k_2}(R)\propto k_2(\cos(2k_2R),\sin(2k_2R))$ of opposite 
vector chirality defined as $\vec{c}=\vec{M}(R)\times \vec{M}(R+dR)$. 
The chirality $\vec{c}_{k_2}= \sin(2k_2dR)\hat{e}_\phi$ points in the $\phi$-direction 
and forms with the directions $\hat{e}_z$ and $\hat{e}_r$ a right handed
coordinate system, while $\vec{c}_{k_1}= -\sin(2k_1dR)\hat{e}_\phi$ points in the 
$-\phi$-direction and forms with $\hat{e}_z$ and $\hat{e}_r$ a left-handed coordinate system. Accordingly,
we call $\vec{M}_{k_1}$ and $\vec{M}_{k_2}$ a left- and right-rotating 
skyrmionic wave, respectively. Our definition of a skyrmion in the actual work 
designates well-defined magnetic waves which are (i) centered around the adatoms,
and have (ii) fixed rotation sense~\cite{roessler}.   
One great virtue of skyrmions is their 
topological nature that is protected under the presence of reasonable
size magnetic fields.

This multi-skyrmionic magnetic waves can be destroyed or manipulated by modifying the 
spin-orbit interaction. By switching it off, only the magnetization along the $z$-direction 
would be finite. One way to tune the Rashba effect is to change the substrate nature 
gradually~\cite{asonen,lindgren}. Such an experiment is very difficult to realize but we propose another tuning 
method: Instead of the Fermi energy, one could probe at different energy varying the bias voltage
between tip and sample. For example, by decreasing the value of the energy probed, the wavelength 
corresponding to $2k$ decreases which is accompanied by a decrease in 
the number of nodes. This is observable in the two additional examples depicted in 
Fig.~\ref{dos_yaxis} calculated at 140~meV and 20~meV. The final multi-skyrmionic 
waves have a different texture but share the common feature in the phase switch at 60~\AA.

One 
should bear in mind that, as for the LDOS, two different regimes have to be 
expected for the magnetization components depending on the sign of $k_2$ (see supplementary material). This 
suggests that the scanning tunneling spectroscopy (STS) experiment of Ast {\it et al.}~\cite{ast}, but spin-polarized, 
would be also a way to verify our predictions. 

\begin{figure}%[ht!]
\begin{center}
\includegraphics*[width=1.\linewidth]{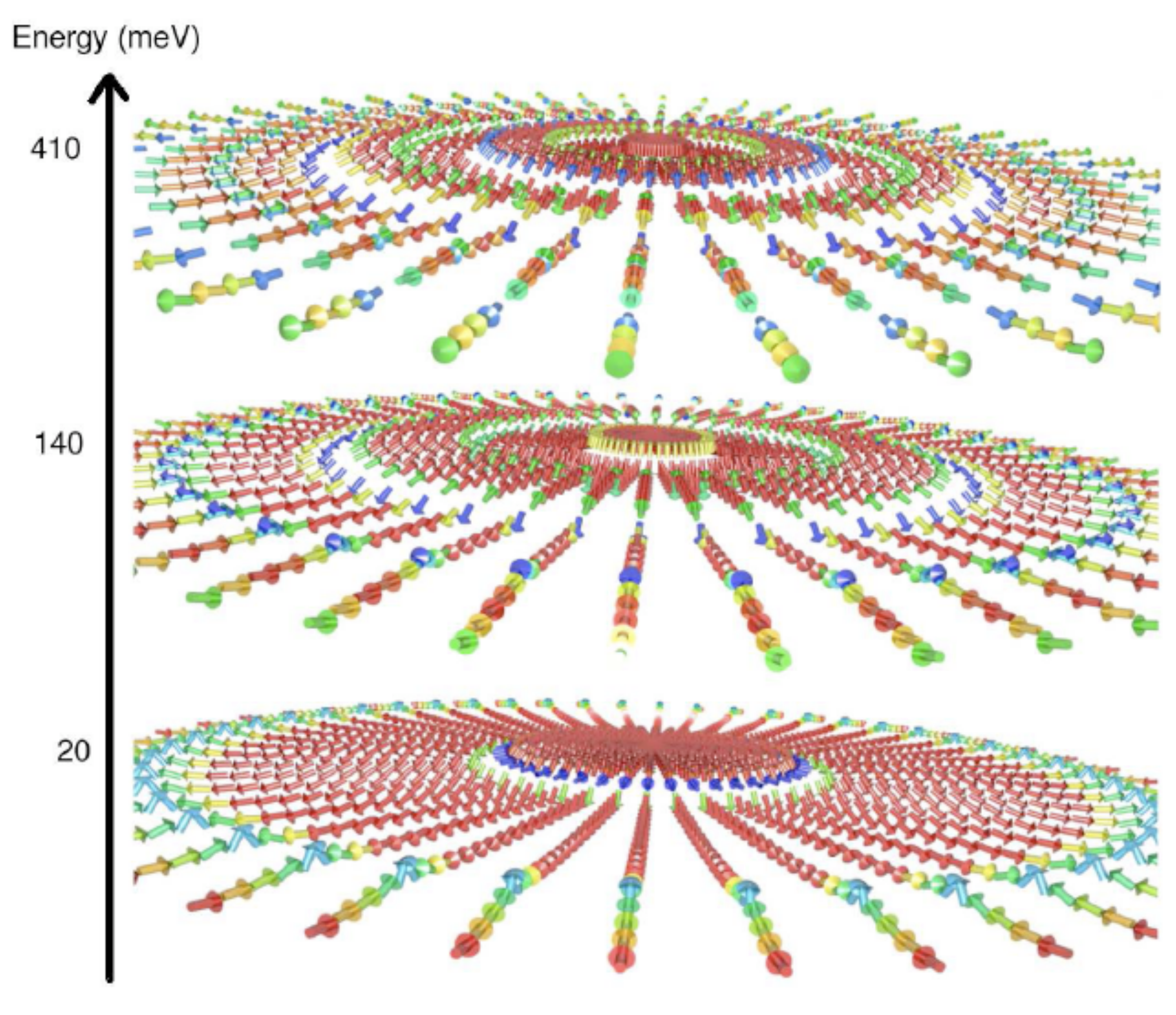}
\end{center}
\caption{Non-collinear spin-configuration of the electron-gas. 
Spectroscopic skyrmionic-like spin-textures of the 
electron-magnetization surrounding the adatoms at 
different energies (410~meV, 140~meV and 20~meV). The number of nodes 
diminishes when decreasing the energy leading to smoother rotating 
spherical magnetic waves.}
\label{spectroscopic_skyrmion}
\end{figure}

To summarize, we have revealed a new type of spin-texture (see Fig.~\ref{spectroscopic_skyrmion}) induced 
in a confined electron gas subject to the Rashba effect by 
a magnetic atom. These structures can be understood as a kind of combination of skyrmionic-like waves of opposite
chirality. 
By tuning the energy level or the spin-orbit strength, 
the observation of the final spin-texture is 
possible because of the large magnitude of the induced magnetization and the creation and exploration of 
new complex configurations that  manageable 
with state of the art experiments. Such non-trivial magnetic Friedel oscillations have an impact on the magnetic interactions between adatoms 
and other nanostructures and consequently on their magnetic behavior. 
 Simple asymptotic expressions for the induced magnetizations are derived which offer 
a simple understanding of new types of experiments involving the manipulation and construction of adatom based spin-nanostructures.

We thank Prof.~K.~Scharnberg for stimulating discussions 
at the initial stage of the work, and Drs.~P.~Lazic and   
H.~Schumacher for their assistance in preparing some of the figures. 
SL acknowledges the support of the HGF-YIG Programme VH-NG-717 (Functional nanoscale structure and probe simulation laboratory, Funsilab).

\end{document}